\documentclass[fleqn,twoside]{article}
\usepackage{espcrc2}
\usepackage{graphicx} 
\usepackage{latexsym}

\newcommand{\AmS}{{\protect\the\textfont2
  A\kern-.1667em\lower.5ex\hbox{M}\kern-.125emS}}

\title{One Loop Renormalization of Fermilab Fermions}

\author{Matthew A. Nobes\address[SFU]{Department of Physics, 
        Simon Fraser University, Burnaby, British Columbia, Canada}
         and
        Howard Trottier\addressmark[SFU]}
       
\begin{document}

\begin{abstract}
We discuss the current status of our automatic perturbation theory program as applied to
Fermilab Fermions.  We give an overview of our methods, a discussion of tree level
matching, and one loop results for the coefficients of the higher dimension 
kinetic operators.
\end{abstract}

\maketitle

A dominant source of error in current lattice calculations is the errors due to finite lattice
spacing $a$.  One way to reduce these errors is to use improved actions.
The general structure for an improved action is
\begin{equation}
\mathcal{L} = \mathcal{L}_{0} + \sum_{n=0}^{\infty} c_{n}(g_{0},m_{0}) \mathcal{L}_{n}
\end{equation}
The series here is an expansion in the dimensionality of the various
operators.  Each of these terms comes with a new coupling constant
$c_{n}$.  

In order to use this type of action we must do two things, we must truncate the series
at some specified order in $n$ and we must calculate the new couplings somehow. 
Fixing the new couplings can be done in a number of ways, 
for example,  the first step could be to make an expansion in powers of $m_{0}$ for light quarks,
or powers of $1/m_{0}$ for heavy quarks.  The remaining dependence on $g_{0}$ can be determined
perturbativly.

It is desirable to have a method of improvement which does not rely on light or heavy
quark mass expansions.  One such approach is the Fermilab approach \cite{mainfermi} which  
orders the expansion operators by dimension only (this amounts to a small momentum 
expansion).  The unimproved fermilab action consists of dimension two and four operators,
\begin{eqnarray}
S_{0} & = & \int dx \bar{\psi} \left\lbrace m_{0} + \frac{1+\gamma_{0}}{2} D^{-}_{0} -
 \frac{1-\gamma_{0}}{2}D^{+}_{0} \right. \nonumber \\
& + & \left. \zeta \mathbf{\gamma}\cdot\mathbf{D} -\frac{r_{s}\zeta}{2} \mathbf{\triangle} \right\rbrace  \psi  
\end{eqnarray}
For the definitions of the various derivatives we refer the reader to \cite{mainfermi}.  This action includes
an additional redundant dimension five operator, whose coefficient $r_{s}\zeta$ can be tuned to remove
the fermion doubling problem. 

To improve to $\mathcal{O}(a^{2})$ a large number of of dimension five and six operators must
be added.  The identification and tree level matching of these operators is
discussed in \cite{bugra}, we will not repeat it here.   
This report is concerned primarily with the one loop determination of the coefficients of the
dimension six kinetic energy operators,
$ \lbrace \mathbf{\gamma} \cdot \mathbf{D} , D^{2} \rbrace$ and 
$ \gamma_{i} D_{i}^{3}$.  For this matching we will set all of the other improvement
coefficients (again see \cite{bugra} for a full list) to
zero.  This leaves the action,
\begin{equation} \label{action}
S  = S_{0} + \int dx \bar{\psi} \left[ c_{1} \lbrace \mathbf{\gamma} \cdot \mathbf{D} , D^{2} \rbrace
 +  c_{2} \gamma_{i} D_{i}^{3} \right] \psi.
\end{equation}
For these kinetic operators the determination of the coefficients is
very straightforward.  For any lattice action, the quark 
energy can always be expanded in powers of the three momentum
\begin{equation} \label{fullenergy}
E(\mathbf{p}) = M_{1} + \frac{\mathbf{p}^{2}}{2M_{2}} - \frac{w_{4}}{6} \sum_{i}p_{i}^{4}
- \frac{(\mathbf{p}^{2})^{2}}{8M_{4}^{3}} + \mathcal{O}(p^{6})
\end{equation}
By tuning the three couplings $\zeta$, $c_{1}$ and $c_{2}$, we can imposed various conditions
on this action.  The coupling $\zeta$ can be tuned to ensure that $M_{1}=M_{2}$, $c_{1}$ and $c_{2}$
can be tuned to ensure that $M_{2}=M_{4}$ and $w_{4}=0$.  The latter condition restores
rotational symmetry to $\mathcal{O}(a^{2})$.  Because it only produces an 
overall shift in the zero of energy it is not necessary to 
tune $M_{1}=M_{2}$ using $\zeta$.  Rather, we'll
just set $\zeta=1$ for convenience.

With $\zeta=1$ we have two remaining improvement conditions, $M_{2}=M_{4}$ and $w_{4}=0$.
We impose these in perturbation theory.  The tree level calculation has
been preformed in \cite{bugra}.  Setting $r_{s}=\zeta=1$, the tree level values are
\begin{eqnarray} \label{treec1}
c^{(0)}_{1}(m_{0}) & = & \frac{m_{0}(2+m_{0})}{32} \nonumber \\
& \times & \left[\left(\frac{1}{1+m_{0}}+\frac{2}{m_{0}(2+m_{0})}\right)^{3}\right. \nonumber \\
& - & \frac{1}{(1+m_{0})^{2}} - \frac{8(1+m_{0})}{m_{0}^{2}(2+m_{0})^{2}} \nonumber \\
& - & \left. \frac{8}{m_{0}^{3}(2+m_{0})^{3}} - \frac{4}{m_{0}^{2}(2+m_{0})^{2}} \right],
\end{eqnarray}
and,
\begin{eqnarray} \label{treec2}
c^{(0)}_{2}(m_{0}) & = & \frac{-m_{0}(2+m_{0})}{12} \nonumber \\
& \times & \left[\frac{1}{4(1+m_{0})}+\frac{2}{m_{0}(2+m_{0})}\right].
\end{eqnarray}
In keeping with the Fermilab approach, these coefficients have their full dependence on the 
bare mass.  Notationally, bracketed superscripts on any quantity denote 
the order of the expansion in the bare coupling.  For example, the rest mass $M_{1}$ has
the expansion
\begin{equation}
M_{1} = M_{1}^{(0)} + g_{0}^{2} M_{1}^{(2)} + g_{0}^{4} M_{1}^{(4)} + \ldots
\end{equation}

The one loop calculations detailed below have been preformed using our automatic 
perturbation theory methods.  The core of this method is the 
use of the L\"uscher -- Weisz vertex generation algorithm \cite{luscher}, with
some straightforward modifications to include arbitrary quark actions.
The vertex rules generated by this method can be used to construct Feynman diagrams, and VEGAS
can be used to preform the loop sums.  Additionally we use triple twisted periodic boundary 
conditions, with $L=200$, as an infrared regulator.
Further details on how these types of calculations
are preformed
can be found in \cite{nobes}.

\begin{figure}[t]
\begin{center}
\input{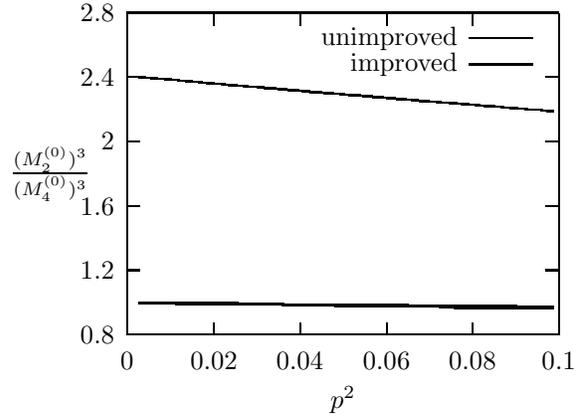}
\end{center}
\vspace{-35pt}
\caption{Tree level check on the energy, for a bare mass of one.  The points show the 
ratio $M_{2}^{(0)}/M_{4}^{(0)}$ over a range of momentum.  The line 
labeled unimproved was generated with the bare action,
the line labeled improved was generated with $c_{1}^{(0)}$
and $c_{2}^{(0)}$ equal to the tree level values, 
(\ref{treec1}) and (\ref{treec2})}
\vspace{-15pt}
\label{fig:treenergy}
\end{figure}

The automatically generated vertex rules can be used to check the tree level calculations 
in \cite{bugra}.  Figure \ref{fig:treenergy} shows the ratio of masses
extracted from the automatically generated quark propagator
over a range of momenta.  Correct tree level matching of $c_{1}^{(0)}$ and $c_{2}^{(0)}$
should give equation (\ref{fullenergy}) with $M_{2}^{(0)}=M_{4}^{(0)}$ over the full range of momenta.  Clearly
the improved action satisfies this, whereas
the unimproved action does not. 
We have also preformed this tree level checking for the 
temporal one gluon vertex, verifying the expressions for
the coefficients $c_{e}$, $c_{6}$ and $c_{8}$ in 
\cite{bugra}.  Work is currently underway to fix and 
confirm the remaining tree level coefficients.

\begin{figure}[t]
\begin{center}
\input{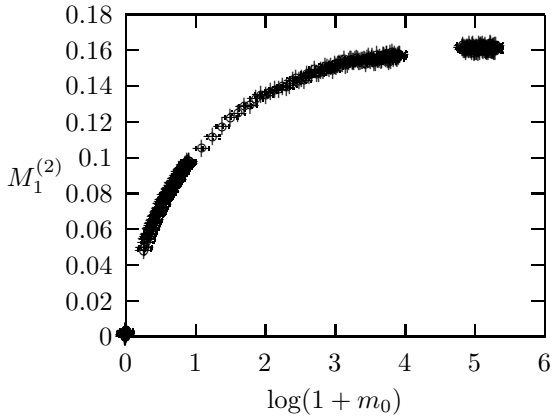}
\end{center}
\vspace{-25pt}
\caption{One loop rest mass, over a range of input bare masses} \label{fig:m12}
\vspace{-20pt}
\end{figure}

We turn now to the one loop calculation of $c_{1}$ and $c_{2}$.
In order to fix these coefficients we must compute
the quark self energy to one loop.  Details of how to preform this calculation can
be found in \cite{mertens}.  Our calculation mirrors \cite{mertens}, apart from the different action 
((\ref{action}) with (\ref{treec1}) and (\ref{treec2})), and
the method of obtaining the Feynman rules.
Expanding (\ref{fullenergy}) in
powers of the bare coupling gives the one loop energy (recall $M_2^{(0)}=M_{4}^{(0)}$)
\begin{eqnarray} \label{e2}
E^{(2)} & = & M_{1}^{(2)} - \mathbf{p}^{2} \frac{M_{2}^{(2)}}{2\left(M_{2}^{(0)}\right)^{2}} \nonumber \\
 & - & w_{4}^{(2)} \sum_{i} p_{i}^{4} + (\mathbf{p}^{2})^{2} 
\frac{3M_{4}^{(2)}}{8\left(M_{2}^{(0)}\right)^{2}}.
\end{eqnarray}
In terms of the quantities that appear here, the improvement conditions are,
\begin{equation}\label{impcon}
M_{2}^{(2)} = M_{4}^{(2)}, \quad w_{4}^{(2)}=0.
\end{equation}

The one loop rest mass $M_{1}^{(2)}$ is  easy to evaluate numerically.  Figure \ref{fig:m12}
shows its value over a wide range of bare masses.  The limit of large bare
masses is approaching the value of $M_{1}^{(2)} = 0.1681(4)$ reported in \cite{mertens}.

\begin{figure}[!h]
\begin{center}
\input{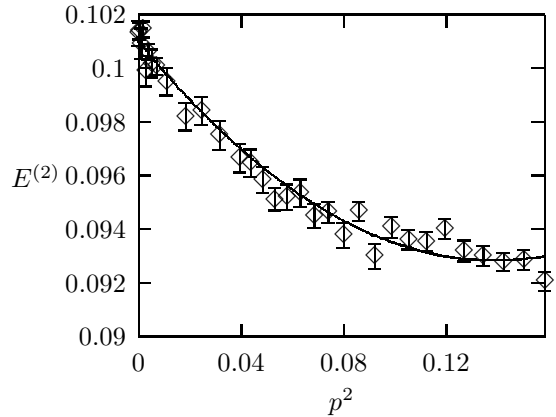}
\end{center}
\vspace{-25pt}
\caption{$\bar{E}^{(2)}$ over a range of $p_{1}$, the line is the best fit}
\vspace{-20pt}
\label{fig:ebarx}
\end{figure}

To compute $c_{1}$ and $c_{2}$ we
start by computing $E^{(2)}$ using the tree level coefficients.  Quantities computed with only the tree
level values for $c_{1}$ and $c_{2}$ will be denoted with bars.  Setting $p_{y}=p_{z}=0$, and
computing $\bar{E}^{(2)}$ over a range of (small) $p_{x}$ allows $\bar{M}_{2}^{(2)}$ and the
combination 
$\bar{\beta} = \frac{3\bar{M}_{4}^{(2)}}{8\left(M_{2}^{(0)}\right)} - \bar{w}_{4}^{(2)}$
to be extracted from fits to the data.  To separate $\bar{M}_{4}^{(2)}$ from
$\bar{w}_{4}^{(2)}$ we compute $\bar{E}^{(2)}$ at fixed $p_{x}$ over a range of
$p_{y}$.  The cross term $p_{x}^{2}p_{y}^{2}$ has coefficient $\frac{3\bar{M}_{4}^{(2)}}{4\left(M_{2}^{(0)}\right)}$ which
can be extracted from a fit to the numerical values of the one loop energy, as illustrated in Figure \ref{fig:ebarx}.
This procedure gives the following values ($m_{0}=1$),
\begin{eqnarray}
M_{1}^{(2)}  =  0.08561(60) & \quad & \bar{M}_{2}^{(2)}  =  0.234(10) \nonumber \\
\bar{M}_{4}^{(2)}  = 0.263(47)  & \quad & \bar{w}_{4}^{(2)}  =  -0.241(52) \nonumber
\end{eqnarray}

The coefficients $c_1^{(2)}$ and $c_2^{(2)}$ provide the counterterms
that are necessary to correct for these one-loop contributions to the energy,
in order to satisfy the improvement conditions (\ref{impcon})
Hence we obtain the preliminary values (recall $m_{0}=1$):
\begin{equation}
c_1^{(2)}  =  0.22(20) \quad c_2^{(2)} =  -2.20(56).
\end{equation} 
These values ensure the matching conditions are satisfied and
give the value
$M_{2}^{(2)}=M_{4}^{(2)}=0.232(10)$.

To conclude, this report illustrates the application of automatic perturbation theory to
the Fermilab fermion action.  These techniques can be used, either to do or
to check, tree level matching, as well as one loop matching.  We have 
presented results for the one loop rest mass $M_{1}^{(2)}$ along with a first
matching calculation for $c_{1}$ and $c_{2}$.  Work is now underway 
to extend
the one loop matching presented here to more values of the bare mass at higher precision, and
more coefficients.

We thank A. Kronfeld, A. El-Khadra and P. Mackenzie 
for fruitful discussion. HDT is supported by NSERC.

\end{document}